\documentstyle[12pt]{article}
\renewcommand{\baselinestretch}{1}
\begin{document}
\title{Are "Penguins" Black-and-White?}
\author{P. \.{Z}enczykowski\\
\\
Departement of Theoretical Physics\\
The H. Niewodnicza\'nski Institute of Nuclear Physics\\
Radzikowskiego 152, Krak\'ow, Poland\\
}
\maketitle
\begin{abstract}
Contributions of low-energy "eye" and "figure-eight" quark diagrams
to the $K \rightarrow \pi$ weak transitions are studied in a hadron-
level phenomenological approach.  It is shown that
these contributions may be estimated by considering
meson-cloud effects.
If all intermediate mesons under consideration are degenerate
only the "eye" (low-energy penguin) diagram is nonvanishing. When
allowance is made for smaller mass of pseudoscalar mesons, the
contribution of "figure-eight" diagrams turns out to enhance
the $\Delta I = \frac{1}{2}$ (suppress the $\Delta I = \frac{3}{2}$)
amplitudes naturally. The overall long-distance-induced enhancement
of the ratio of the
$\Delta I = \frac{1}{2}$ amplitudes over the $\Delta I =
\frac{3}{2}$ amplitudes is estimated at around 4-8.
\\

\end{abstract}

\newpage
\baselineskip = 10 pt
\renewcommand{\baselinestretch}{1.685}
\small
\normalsize

\section{Introduction}

After almost 40 years since the discovery
of the $\Delta I = 1/2$ rule in strangeness-changing weak
hadronic decays, its origin still eludes our understanding
(for a recent review see ref.\cite{ChengDeltaI}).
Dominance of the $\Delta I = 1/2$ amplitudes over those
with $\Delta I = 3/2$ requires a significant enhancement of
the former and/or suppression of the latter.
While for nonleptonic baryon decays at least part of the effect
stems from the Pati-Woo
theorem \cite{PatiWoo}, according to which
the symmetry of baryon wave functions ensures vanishing of
the $\Delta I =3/2$ amplitude,
no such symmetry-based mechanism is available for kaon decays.

The required effects can be obtained to some extent from
perturbative QCD. Short-distance QCD corrections
modify the effective weak Hamiltonian and lead to an enhancement
of the $\Delta I =1/2$ (suppression of the $\Delta I = 3/2$)
operators \cite{Gaillard}.
In addition, a new purely $\Delta I =1/2$ mechanism - the so-called
penguin operator - appears. Its contributions add
constructively to those of standard $\Delta I = 1/2$ operators.
Detailed studies \cite{Peng} show, however,
that the original claim of a large penguin contribution is incorrect.
This contribution remains small even if one takes into account the
increase, over the value quoted in ref.\cite{GilmWise}, of the
real part of the penguin Wilson coefficient due to the
incomplete GIM cancellation above the charm quark mass \cite{BBG}.

Dropping the so-called Fierz contributions (which has been argued to
be justified in the 1/N expansion, \cite{BurasGerardRuckl}) does help
a little, but a large discrepancy still remains \cite{ChengDeltaI}.
In fact, for consistency with the spirit of the $1/N$ expansion, the
Fierz terms should be considered along with nonfactorizable terms
of the same order. Starting from an effective chiral Lagrangian, such
subleading $1/N$ contributions have been calculated
in ref.\cite{BBG1/N} as nonfactorizable pseudoscalar meson
loop corrections to $K \rightarrow 2 \pi$.
Their contribution has been found to be
of the same order as that of the genuine factorizable terms.
In ref.\cite{ChengDeltaI} the following effects are mentioned
as contributing to the subleading terms of the $1/N$ approach:
the Fierz-transformed contributions, final state-interactions,
low energy "eye" graphs, and soft gluon exchanges
between two quark loops in "figure-eight" graphs.
Because of the long-distance nature of the last three mechanisms,
their evaluation from the first principles of QCD is possible
on the lattice only. In practice it is
the $K \rightarrow \pi$
matrix elements that are more amenable to such calculations.
{}From these, the
$K \rightarrow 2\pi$ amplitudes are then obtained
by means of current algebra.
Within very large statistical and systematic uncertainties the
lattice calculations\cite{Lattice}
support the $\Delta I =1/2$ enhancement and indicate that the purely
$\Delta I =1/2$ "eye" graphs dominate over the "figure-eight" graphs.

The contribution from the "eye" and "figure-eight" graphs of the
quark-level description must be contained in the
meson-cloud (or unitarity) effects of the hadron-level
(as these include all confinement effects,
see also ref.\cite{Kounnas}).  In fact,
it has been found repeatedly by many authors that such meson cloud
effects are very important in many areas of hadron physics,
improving the predictions of
the standard quark model.
For a unitarity-oriented view of hadron
spectroscopy see refs.\cite{Tornqvist,Tornqv1991,Zenczykowski}.
Meson-cloud effects have also been found instrumental in
several other places
\cite{Truong,Szczurek}.
Consideration of their effects in weak nonleptonic hyperon decays
yields an explanation of the deviation of the $f/d$ ratio from
the naive valence quark model value of -1 to its observed values
of about -2 \cite{Zen92}.
It is therefore of great interest to perform
a similar phenomenological analysis of meson cloud effects
in $K \rightarrow 2\pi$ decays to see whether and how they may help
to explaine relative sizes of the relevant $\Delta I = 1/2$
and $\Delta I =3/2$ amplitudes.
In this paper an analysis of this type is carried out.
We study
the $K \rightarrow \pi$
transition matrix elements and
show in detail how hadron-level
effects from various two-meson intermediate states
contributing to these transitions build up the "eye" (low-energy
penguin) and the "figure-eight" diagrams of the quark level.
An estimate of the relative and absolute sizes of the "eye" and
"figure-eight" diagrams
is also given.

\section{Hadronic loop contributions to the $K \rightarrow \pi$
transitions}
The effect of pseudoscalar meson loop contributions to
$K \rightarrow
2 \pi$ was studied
in dispersion relation framework \cite{Truong}, and in
chiral approach \cite{BBG1/N,BijGubBelkov}.
In more phenomenological way such meson rescattering FSI
effects are discussed in ref.\cite{FSI}.
In this paper we are concerned with meson loop (hadron sea)
effects in $K \rightarrow \pi$ transitions themselves (see Fig.1).
If only ground-state mesons are permitted in the loop, at least one
of them must be a vector meson  (the allowed intermediate states
are PV+VP and VV
(P-pseudoscalar, V-vector mesons). Although all these two-particle
states are much heavier than the PP ones that were considered in
refs.\cite{BBG1/N,Truong,BijGubBelkov,FSI},
their contribution
is expected to be significant as evidenced by estimates of their
effects in hadron spectroscopy\cite{Tornqvist,Zenczykowski}.
A transparent way to include both pseudoscalar and vector mesons
in the intermediate state
is to use general ideas of the unitarized quark model
of ref.\cite{Tornqvist}.

What we want to estimate here is, in essence,
the contribution from
virtual two-meson continuum states admixed
into the wave functions of the standard quark model.
We shall disregard the virtual states composed of charmed mesons
as such states lie much higher (by about 2 GeV) than those
built of light flavours.
In the approach of ref.\cite{Tornqvist} the admixture probability
$|c_{M_{1}M_{2}}|^{2}$ of the $\mid M_{1}M_{2} >$ two-particle state
relative to the "pure" quark-model
state for meson M is given by
\cite{OnoTorn}
\begin{equation}
\label{eq:mixsize}
|c_{M_{1}M_{2}}|^{2} = L(M\rightarrow M_{1}M_{2})
 [Tr(F_{M}^{\dagger}F_{M_{1}}F_{M_{2}})+
C_{M}C_{M_{1}}C_{M_{2}}Tr(F_{M}^{\dagger}F_{M_{2}} F_{M_{1}})]^{2}
\end{equation}
where, for ground-state mesons $M_{1}M_{2}$, we have
\begin{eqnarray}
\label{eq:mixground}
L(M\rightarrow M_{1}M_{2})&=&
S (M \rightarrow M_{1}M_{2})~ I \nonumber\\
& \equiv &
S (M \rightarrow M_{1}M_{2})~ \frac{1}{\pi} \frac{f^{2}}{\pi}
\int_{thr}^{\infty}
\frac{\frac{k^{3}}{\sqrt{s}} \exp{-(\frac{k}{k_{cutoff}})^{2}} }
{(m^{2}-s)^{2}} ds
\end{eqnarray}
The trace factor
in Eq.\ref{eq:mixsize}
($F_{M}$ is the SU(3) matrix corresponding to meson M)
gives F- or D- type flavour couplings
depending on the sign of $C_{M}C_{M_{1}}C_{M_{2}}$ (where
$C_{M}$ is the charge conjugation quantum number of meson $M$).
The spin-weight factors $S(M \rightarrow M_{1}M_{2})$
are equal to $\frac{1}{4}$,$\frac{1}{4}$,$\frac{1}{2}$ for
($M, M_{1}M_{2}$) being ($P,PV$), ($P,VP$), ($P,VV$) respectively,
and they sum up to $1$.
The overall size
of the two-meson admixture is fixed by the size of the
coupling constant $f = f_{\rho N N}=5.14$ (Eq.\ref{eq:mixground}),
 and by $k_{cutoff}$ which
is related to the (harmonic oscillator) meson size by
$R^{2}_{M} = \frac{6}{k^{2}_{cutoff}}$.
The size of the integral $I$ depends on the actual positions
of thresholds. As a rough estimate of effects under discussion,
we evaluate
the integral $I$ using
for the external ($M$) meson mass the value
{}~~$m=\frac{1}{2}(m_{\pi}+m_{K})=0.32~GeV $~~ for
two sets of masses of intermediate mesons:
1) the degenerate case with $m_{V}=m_{P}=0.9~GeV$,
and 2) the light pseudoscalar meson case with
$m_{V}=0.9~GeV$, $m_{P}\approx m=
0.32~GeV$. The obtained values are gathered in Table 1 for
$k_{cutoff}=
0.6,0.7,0.8~GeV$ ($R_{M}=0.80,0.69,0.60~fm$).
In the unitarized quark model of ref.\cite{Tornqvist,Tornqv1991}
the value of $k_{cutoff}=0.7~GeV$ gives the best description of
meson spectra.

Since, according to Eq.\ref{eq:mixsize}, admixtures of two-meson
$|~\rho \pi~>$, $|~ \rho~\eta~>$, etc. states to $\pi$ meson
($|~\rho~K~>$ etc. to $K$)
are all to be considered, we will have to deal with the
$K \rightarrow \eta $ transitions as well.
With the Fierz terms dropped and small short-distance
penguin contributions neglected,
standard QCD-corrected short-distance calculations give
the following predictions
for the amplitudes:
\begin{eqnarray}
\label{eq:shortdistQCD}
<\pi ^{+} \mid H_{w} \mid K^{+} >& = &
[c_{1}-(c_{2}+c_{3}+c_{4})] ~ X \nonumber\\
<\pi ^{0} \mid H_{w} \mid K^{0} >& = &
\frac{1}{\sqrt{2}}[c_{1}-(c_{2}+c_{3}-2c_{4})] ~ X \nonumber\\
<\eta _{8} \mid H_{w} \mid K^{0} >&=&
\frac{1}{\sqrt{6}}[c_{1}-c_{2}+9c_{3}] ~X \nonumber\\
<\eta _{1} \mid H_{w} \mid K^{0} >&=&
\frac{1}{\sqrt{3}}[c_{1}+5c_{2}] ~X
\end{eqnarray}
where $c_{i}$ are Wilson coefficients
and
\begin{equation}
\label{eq:X}
X =
<\pi^{+} \mid -(d\bar{u}) \mid 0 > < 0 \mid (u\bar{s}) \mid K^{+} >
\end{equation}
with the notation
\begin{equation}
\label{eq:q1q2}
(q_{1}\bar{q}_{2}) \equiv \bar{q}_{2}\gamma_{\mu}(1-\gamma_{5})q_{2}
\end{equation}

Let us express the matrix elements of the parity conserving part of
weak Hamiltonian between pseudoscalar meson states through amplitudes
of definite isospin:
\begin{eqnarray}
\label{eq:isospinampl}
<\pi ^{+} \mid H_{w} \mid K^{+} >& =& \sqrt{\frac{2}{3}}
A_{\frac{1}{2}}-\frac{1}{\sqrt{3}}A_{\frac{3}{2}}\nonumber\\
<\pi ^{0} \mid H_{w} \mid K^{0} >& =& \frac{1}{\sqrt{3}}
A_{\frac{1}{2}} + \sqrt{\frac{2}{3}}A_{\frac{3}{2}}\nonumber\\
< \eta_{8} \mid H_{w} \mid K^{0} >& =& B \nonumber\\
< \eta_{1} \mid H_{w} \mid K^{0} >& =& C
\end{eqnarray}

Using the Gilman-Wise values
\cite{GilmWise}:
\begin{eqnarray}
\label{eq:Ccoeff}
c_{1}&=&-2.11 \nonumber\\
c_{2}&=&0.12 \nonumber\\
c_{3}&=&0.09 \nonumber\\
c_{4}&=&0.45
\end{eqnarray}
for the Wilson coefficients, we obtain
from Eqs.\ref{eq:shortdistQCD},\ref{eq:isospinampl}
\begin{eqnarray}
\label{eq:QCDnoFierz}
\frac{A_{\frac{3}{2}}}{A_{\frac{1}{2}}}& =& -0.28 \nonumber\\
\frac{B}{A_{\frac{1}{2}}} & = & 0.20 \nonumber\\
\frac{C}{A_{\frac{1}{2}}} & = & 0.31
\end{eqnarray}
The experimental value for $\left| \frac{A_{\frac{1}{2}}}
{A_{\frac{3}{2}}}\right|$ is around 22, six times larger than
the theoretical value from Eq.\ref{eq:QCDnoFierz}
($\left| \frac{A_{\frac{1}{2}}}{A_{\frac{3}{2}}}\right| $ = 3.6).
When short-distance penguin contribution is included
(with $c_{5} \approx -0.06$) one obtains \cite{ChengDeltaI}
$|A_{\frac{1}{2}}/A_{\frac{3}{2}}|=4.3$, i.e.
\begin{equation}
\label{eq:shdistpeng}
\left( \frac{A_{\frac{1}{2}}}{A_{\frac{3}{2}}}\right) _{out} =
1.2 \left( \frac{A_{\frac{1}{2}}}{A_{\frac{3}{2}}}\right) _{fact}
\end{equation}
an enhancement factor of 1.2 only.
The remaining discrepancy by a factor of around 5
constitutes the $\Delta I = \frac{1}{2}$ puzzle.

 The hadron-sea generated corrections to the matrix elements
of Eq.\ref{eq:isospinampl}
are due to weak Hamiltonian acting in one of $M_{1}$,
$M_{2}$ mesons.
Let the meson in which such a transition
occurs be labelled $M_{1}$ (see Fig.1).
We restrict our considerations to the case when $M_{1}$ is in
the ground state (i.e. $M_{1} = P,V$).
For the sake of our discussion this should be a reasonable
approximation: Quark-antiquark annihilation into a $W$-boson
is expected weaker for excited mesons. Moreover,
the additional contributions arising from weak transition
in an intermediate excited meson should (when estimated along
lines similar to those presented in this paper)
only increase the enhancement/suppression effects
herein discussed (this should become understandable later, after
the discussion of the case $M_{1} = P,V$ ).

On the other hand, both ground-state and excited $M_{2}$ mesons will
be considered. In fact, in strong virtual decays $M \rightarrow
M_{1}M_{2}$ the p-wave
(that must appear somewhere to ensure parity conservation in the
production of $q\bar{q}$-pair out of the vacuum)
may reside either between mesons $M_{1} M_{2}$ or within
meson $M_{2}$. The contributions from these two possibilities
should be comparable. The relative size of the two terms
may be fixed by requiring their mutual cancellation in
Zweig-rule-forbidden strong amplitudes
\cite{Tornqvist}. This relative size may also be obtained under
some additional assumptions through explicit calculations in the
$^{3}P_{0}$-model \cite{Zenunpubl}. The
spin-flavour factors
($ [Tr(F_{M}^{\dagger}F_{M_{1}}F_{M_{2}})+
C_{M}C_{M_{1}}C_{M_{2}}Tr(F_{M}^{\dagger}F_{M_{2}} F_{M_{1}})]^{2}$
*$S(P\rightarrow M_{1}M_{2}$) corresponding
to the total contribution from all possible intermediate states
under consideration are gathered in Table 2.

In the normalization of Eq.\ref{eq:mixsize} the contributions from
the $M_{1}M_{2}$ =
 $\underline{P}V$ two-meson states (the meson
undergoing weak transition underlined for clarity) are easily
calculable to be:
\begin{eqnarray}
\label{eq:full1a}
A_{\frac{3}{2},loop}& =& - 2 L(P\rightarrow PV) A_{\frac{3}{2}}
\nonumber\\
({\bf 27})~~~~~~~~~~~
\frac{1}{\sqrt{10}}(A_{\frac{1}{2},loop}-3B_{loop})&=&
- 2 L(P\rightarrow PV) \frac{1}{\sqrt{10}}(A_{\frac{1}{2}}-3B)
\phantom{xxxxxx}
\nonumber\\
({\bf 8})~~~~~~~~~~~~
\frac{1}{\sqrt{10}}(3A_{\frac{1}{2},loop}+B_{loop})&=&
+ 3 L(P\rightarrow PV) \frac{1}{\sqrt{10}}(3A_{\frac{1}{2}}+B)
\nonumber\\ C_{loop}& =& 0
\end{eqnarray}
(Where such an assignment is not obvious, the
SU(3) classification of the amplitude is given on the left.)
When the p-wave excitation resides in the $M_{2}$ meson, total
contribution from the S- and D- wave two-meson states
$\underline{P}V^{*}$ ($V^{*}$ =$ S$(scalar, $J^{PC}=0^{++}$),
$A$(axial, $1^{++}$),
$T$(tensor,$2^{++}$) mesons) is:
\begin{eqnarray}
\label{eq:full1b}
A_{\frac{3}{2},loop}&=&+2 L(P\rightarrow PV^{*}) A_{\frac{3}{2}}
\\
\frac{1}{\sqrt{10}}(A_{\frac{1}{2},loop}-3B_{loop})&=&
+ 2 L(P\rightarrow PV^{*})\frac{1}{\sqrt{10}}(A_{\frac{1}{2}}-3B)
\nonumber\\
\frac{1}{\sqrt{10}}(3A_{\frac{1}{2},loop}+B_{loop})&=&
+ \frac{1}{3} L(P\rightarrow PV^{*})
(\frac{1}{\sqrt{10}}
\left( 3A_{\frac{1}{2}}+B) - 4\sqrt{5}C \right)  \nonumber\\
 C_{loop}& =&
+ \frac{1}{3} L(P\rightarrow PV^{*})
\left( - 4 \sqrt{5}\frac{1}{\sqrt{10}}
(3A_{\frac{1}{2}}+B) + 8 C\right)
\phantom{xxx}
           \nonumber
\end{eqnarray}
In writing Eq.\ref{eq:full1b} we summed the contributions from the S-
and D-waves by assuming that they are equal apart from their
difference in weight
(see Table 1). This should be a reasonable assumption
since, at small values of $m$ ($\approx m_{\pi}$ or $m_{K}$), we are
away from thresholds where such differences might be important.
In the $^{3}P_{0}$ model, factors $L(P\rightarrow PV^{*})$ are
given by a formula similar to Eq.\ref{eq:mixground}.

Meson $M_{1}$ need not be a pseudoscalar meson.
It may be a vector meson as well.
For weak transitions in vector mesons we introduce
notation analogous to that of Eq.\ref{eq:isospinampl}:
the $K^{*} \rightarrow \rho$ transitions are described by
amplitudes
$A^{V}_{\frac{1}{2}}$,$A^{V}_{\frac{3}{2}}$
of definite isospin etc.
When $M_{1}=V$ we have contributions from $\underline{V}P$ and
$\underline{V}P^{*}$ ($P^{*} = B$ (axial $J^{PC}=1^{+-}$)
meson) two-meson states. They are, respectively:
\\
a) for $\underline{V}P$ loops:
\begin{eqnarray}
\label{eq:full2a}
A_{\frac{3}{2},loop}&=&- 2 L(P\rightarrow VP) A^{V}_{\frac{3}{2}}
\nonumber\\
\frac{1}{\sqrt{10}}(A_{\frac{1}{2},loop}-3B_{loop})&=& - 2
L(P\rightarrow VP) \frac{1}{\sqrt{10}}(A^{V}_{\frac{1}{2}}-3B^{V})
\nonumber\\
\frac{1}{\sqrt{10}}(3A_{\frac{1}{2},loop}+B_{loop})&=& + 3
L(P\rightarrow PV) \frac{1}{\sqrt{10}}(3A^{V}_{\frac{1}{2}}+B^{V})
\nonumber\\ C_{loop}& =& 0
\end{eqnarray}
b) for $\underline{V}P^{*}$ loops:
\begin{eqnarray}
\label{eq:full2b}
A_{\frac{3}{2},loop}&=&+2 L(P\rightarrow VP^{*}) A^{V}_{\frac{3}{2}}
\nonumber\\
\frac{1}{\sqrt{10}}(A_{\frac{1}{2},loop}-3B_{loop})&=& + 2
L(P\rightarrow VP^{*})\frac{1}{\sqrt{10}}(A^{V}_{\frac{1}{2}}-3B^{V})
\nonumber\\
\frac{1}{\sqrt{10}}(3A_{\frac{1}{2},loop}+B_{loop})&=&
+ \frac{1}{3} L(P\rightarrow VP^{*})
\left( \frac{1}{\sqrt{10}}
(3A^{V}_{\frac{1}{2}}+B^{V}) - 4\sqrt{5}C^{V}\right)  \nonumber\\
 C_{loop}& =&
+ \frac{1}{3} L(P\rightarrow VP^{*})
\left( - 4 \sqrt{5}\frac{1}{\sqrt{10}}
(3A^{V}_{\frac{1}{2}}+B^{V}) + 8 C^{V}\right)
\nonumber\\
&&
\end{eqnarray}
Finally, contributions from the $\underline{V}V$ and
$\underline{V}V^{*}$ diagrams are:

a) for the $\underline{V}V$ loops:
\begin{eqnarray}
\label{eq:full3a}
A_{\frac{3}{2},loop}&=& + 2 L(P\rightarrow VV) A^{V}_{\frac{3}{2}}
  \nonumber\\
\frac{1}{\sqrt{10}}(A_{\frac{1}{2},loop}-3B_{loop})&=& + 2
L(P\rightarrow VV)\frac{1}{\sqrt{10}}(A^{V}_{\frac{1}{2}}-3B^{V})
  \nonumber\\
\frac{1}{\sqrt{10}}(3A_{\frac{1}{2},loop}+B_{loop})&=&
+ \frac{1}{3} L(P\rightarrow VV)
\left( \frac{1}{\sqrt{10}}
(3A^{V}_{\frac{1}{2}}+B^{V}) - 4\sqrt{5}C^{V}\right)  \nonumber\\
 C_{loop}& =&
+ \frac{1}{3} L(P\rightarrow VV)
\left( - 4 \sqrt{5}\frac{1}{\sqrt{10}}
(3A^{V}_{\frac{1}{2}}+B^{V}) + 8 C^{V}\right)
  \nonumber\\
&&
\end{eqnarray}

b) for the $\underline{V}V^{*}$ loops:
\begin{eqnarray}
\label{eq:full3b}
A_{\frac{3}{2},loop}&=& - 2 L(P\rightarrow VV^{*}) A^{V}_{\frac{3}{2}}
\nonumber\\
\frac{1}{\sqrt{10}}(A_{\frac{1}{2},loop}-3B_{loop})&=& - 2
L(P\rightarrow VV^{*})
\frac{1}{\sqrt{10}}(A^{V}_{\frac{1}{2}}-3B^{V})
\nonumber\\
\frac{1}{\sqrt{10}}(3A_{\frac{1}{2},loop}+B_{loop})&=& + 3
L(P\rightarrow VV^{*})
\frac{1}{\sqrt{10}}(3A^{V}_{\frac{1}{2}}+B^{V})
\nonumber\\
C_{loop} &=& 0   \nonumber\\
&&
\end{eqnarray}
As already discussed, the relative size of contributions from $M_{2}$
 = $P$,$V$ and $M_{2}$=$P^{*}$,$V^{*}$ is fixed when the validity of
Zweig's rule is ensured by cancellation of contributions from
intermediate states involving mesons of opposite C-parity.
This amounts to putting
$L(P\rightarrow PV^{*}) = L(P\rightarrow PV) =
L(P\rightarrow VP) = L(P\rightarrow V^{*}P) (=\frac{1}{4} I)$
and
$L(P\rightarrow VV^{*}) = L(P\rightarrow VV) (=\frac{1}{2} I)$.
Summing the contributions from all intermediate states considered
we obtain:
\begin{eqnarray}
\label{eq:total}
A_{\frac{3}{2},loop}&=& 0\phantom{xxxxxxxxxxxxxxxxxxxxxxxxxx}
\nonumber\\
\frac{1}{\sqrt{10}}
(A_{\frac{1}{2},loop}-3B_{loop})& =& 0 \nonumber\\
\frac{2}{3\sqrt{10}}(3A_{\frac{1}{2},loop}+B_{loop})
+\frac{\sqrt{5}}{3}C& =& 0  \nonumber\\
\frac{1}{3\sqrt{2}}(3A_{\frac{1}{2},loop}+B_{loop})-
\frac{2}{3}C_{loop}&=& \nonumber\\
=2*6* \frac{1}{4}I
\left\{
\frac{1}{3\sqrt{2}}(3A_{\frac{1}{2}}\right. +
\lefteqn{
B)-\frac{2}{3}C +
3\left( \frac{1}{3\sqrt{2}}(3A^{V}_{\frac{1}{2}}+B^{V})-
\frac{2}{3}C^{V}\right)
\left.\phantom{\frac{|}{|}}\right\} }
& & \nonumber\\
&&
\end{eqnarray}

In the last of equations in (\ref{eq:total}) the overall factor
of "2" on the r.h.s. stems from the fact that weak interaction may
occur in either one of the two intermediate mesons.
{}From
Eqs.(\ref{eq:full1a}-\ref{eq:total})
we see that after summing over the charge-conjugated
$M_{2}=V,V^{*}(P,P^{*})$ meson states, virtual
two-meson states give no contribution to the ~~{\bf 27}-plet $\Delta I
= \frac{1}{2}$ and $\frac{3}{2}$~ transition amplitudes. Furthermore,
only one of the two combinations of octet ($\Delta I=\frac{1}{2}$)
transition amplitudes receives contributions from such states.
We shall estimate the loop contribution to this transition by using
short-distance
QCD-modified factorization approximation
(with Fierz-transformed terms dropped)
for the
$\Delta S =1$
transition
 occurring in meson $M_{1}$.
This gives
\begin{eqnarray}
\label{eq:weakleg}
\frac{1}{3\sqrt{2}}(3A_{\frac{1}{2},loop}+B_{loop})-
\frac{2}{3}C_{loop}& =&
2*6*I*\frac{1}{\sqrt{3}}(c_{1}-5c_{2})
\left[ \frac{X+3X^{V}}{4}\right]
\nonumber\\
&&
\end{eqnarray}
wherein $X^{V}$ is the factorization contribution from weak
transition in intermediate vector meson
\begin{eqnarray}
\label{eq:XV}
X^{V} &=& <\rho^{+} \mid -(d\bar{u}) \mid 0 >
< 0\mid (u\bar{s})\mid K^{*+}>
\nonumber\\
&&
\end{eqnarray}
The loop contribution of Eq.\ref{eq:weakleg} should be compared with
the short distance contribution to this transition amplitude which is
\begin{eqnarray}
\label{eq:shdistcontribution}
\frac{1}{\sqrt{3}}(c_{1}-5c_{2})X
&&
\nonumber\\
&&
\end{eqnarray}
The matrix elements of currents in Eqs.\ref{eq:X},\ref{eq:XV} are
given by
\begin{eqnarray}
\label{eq:decayconstants}
<\pi ^{+} | A^{\mu}|0>& =& f_{\pi}q^{\mu}
\nonumber\\
<\rho ^{+} | V^{\mu}|0>& =& f_{\rho}\varepsilon^{\mu}
\nonumber\\
&&
\end{eqnarray}
where $f_{\pi}=0.13~GeV$, $f_{\rho}=0.17~GeV^{2}$.
In accordance with the $SU(3)$ symmetry used elsewhere in this
paper we assume, for the sake of the order-of-magnitude estimate,
that $f_{K}=f_{\pi}$, $f_{K^{*}}=f_{\rho}$.

Calculation of the $K\rightarrow \pi ,~\eta$
matrix elements
in the vacuum insertion method gives
expressions proportional to the four-momentum squared ($q^{2}$),
(i.e. $<\pi (p) | H^{p.c.}_{W} | K(q)> = p.q~ g_{\pi K}$ etc.)
in agreement with general requirements of chiral symmetry
\cite{SU3,MTT,DGPH,ChengDeltaI}.
Such momentum dependence
is not manifest
in our phenomenological calculations of long-distance effects.
It is well supported by lattice calculations, however
\cite{ChengDeltaI}.
Because of the lack of explicit momentum dependence
there is a problem here
as to what value should be used for the $q^{2}$ of the
pseudoscalar meson undergoing weak transition in the
short-distance factorization contribution with which the loop
effect is compared.
(The contribution from weak transitions in intermediate
pseudoscalar mesons is much smaller than that from corresponding
transitions in vector mesons and, consequently, this ambiguity is
less important in the estimate of loop effects themselves).
As a rough measure we employ
{}~~$q^{2}=\frac{1}{2}(m^{2}_{K}+m^{2}_{\pi})=0.132~GeV^{2}$~~
(see also ref.\cite{DGPH}).
Consequently, the relevant
ratio of factorization contributions $X^{V}$ and $X$ is
\begin{eqnarray}
\label{eq:XVXratio}
\frac{X^{V}}{X}& =&\frac{f^{2}_{\rho}}{f^{2}_{\pi}q^{2}}
\approx 13.0 \nonumber\\
&&
\end{eqnarray}
and the two-meson admixture contributes approximately
\begin{eqnarray}
\label{eq:enhancement}
3~I~\left( 1+3\frac{X^{V}}{X}\right)& \approx & 120~I
\end{eqnarray}
times more than the original factorization contribution.
For $I=0.022$ (from Table 1 for $m_{M_{1}}=m_{M_{2}}=0.9~GeV$)
we obtain an enhancement factor of 2.6.
Clearly,
the bulk of the enhancement obtained comes from
the contribution of weak transitions in intermediate {\em vector}
meson. The hadron-loop-induced enhancement factor of 2.6
should be compared with the standard short-distance estimates of
penguin effects that give a factor of 1.2 (Eq.\ref{eq:shdistpeng}
and ref.\cite{ChengDeltaI}).

\section{Discussion}
Let us see what types of quark-level diagrams are generated by
hadron-level loops under discussion. Consider $\underline{P}V$
and $\underline{P}V^{*}$ intermediate states as an example.
In the contribution from the $\underline{P}V$ loop
(Eq.\ref{eq:full1a}), strong vertices are described by F-type
flavour factors, while for the $\underline{P}V^{*}$ loop
the corresponding couplings are of D-type (see Eq.\ref{eq:mixsize}).
The flavour structure of these strong vertices may be represented
diagrammatically as in Fig.2.  The wavy lines symbolize
confining strong forces.

The structure of the product of flavour factors corresponding
to two strong vertices of the loop is then

a)for $P\rightarrow \underline{P}V \rightarrow P$ loops:
\begin{equation}
\label{eq:Floop}
Tr(F_{M}[F_{M_{1}}^{\dagger},F_{M_{2}}^{\dagger}])
Tr(F_{M'}^{\dagger}[ F_{M'_{1}},F_{M_{2}}])
\end{equation}

b)for $P\rightarrow \underline{P}V^{*} \rightarrow P$ loops:
\begin{equation}
\label{eq:Dloop}
Tr(F_{M}\{F_{M_{1}}^{\dagger},F_{M_{2}}^{\dagger}\})
Tr(F_{M'}^{\dagger}\{ F_{M'_{1}},F_{M_{2}}\})
\end{equation}
Using the equality
{}~~$\sum_{M=1\oplus8}Tr(AM)Tr(AM^{\dagger})=Tr(AB)$,~~
summation over all intermediate mesons $M_{2}$ may be performed,
giving the expression
\begin{eqnarray}
\label{eq:FDloopsummed}
&Tr(F_{M_{1}^{\dagger}} F_{M} F_{M'^{\dagger}} F_{M'_{1}})+
Tr(F_{M} F_{M_{1}^{\dagger}} F_{M'_{1}} F_{M'^{\dagger}})&\nonumber\\
&\mp Tr(F_{M_{1}^{\dagger}} F_{M} F_{M'_{1}} F_{M'^{\dagger}})
\mp Tr(F_{M} F_{M_{1}^{\dagger}} F_{M'^{\dagger}} F_{M'_{1}})&
\nonumber\\
&&
\end{eqnarray}
with $-(+)$ signs for $F(D)$ respectively.
Flavour contractions implicit in the first and the second term of
Eq.\ref{eq:FDloopsummed} are visualised in Fig.3a, while those of
the remaining two terms - in Fig.3b.
The black blob in Fig.1 is replaced in Fig.3 with
boxes marked with dashed lines.
Inside the boxes
the diagrammatic representation of the genuine factorization
prescription is drawn.

Fig.3a represents the familiar low-energy penguin ("eye") diagram,
while Fig.3b is easily recognizable as the "figure-eight"-type
diagram with soft gluon exchanges between two quark loops.
When the internal organization of the weak-interaction box is
taken into account, the "figure-eight" diagram of Fig.3b is
actually equivalent to the $W$-exchange diagram with all possible
soft gluon exchanges between an initial (anti)quark and a final
(anti)quark.
When the summation of two contributions from $M_{2}=V$ and $V^{*}$
(Eq.\ref{eq:FDloopsummed}) is performed with equal weights
(which in the previous section was argued to be a reasonable
approximation),
the "figure-eight" contribution drops out
totally from final formulas (Eq.\ref{eq:total}) and,
consequently, expressions in Eq.\ref{eq:total} correspond to
the low-energy penguin interaction with a {\em u}-quark loop.

In the discussion so far we have assumed that the contributions from
all loops with different internal mesons are essentially identical,
irrespectively of the actual location of the relevant thresholds.
In reality, pseudoscalar mesons
are much lighter than the remaining scalar, axial and tensor mesons.
Consequently, contribution from intermediate states containing a
pseudoscalar meson (especially a pion) will be larger. To see
what effect such nondegeneracy might have, let us assume - as a
very rough approximation - that all pseudoscalar mesons are lighter
than the remaining, still approximately degenerate, vector, axial
and tensor mesons. This idealization corresponds to the expected
dominance of contributions from low-lying thresholds and to small
(and thus negligible)
differences in the overall scale of contributions from the remaining
states.

Using Eqs.(\ref{eq:full1a}-\ref{eq:full3b}) we derive
the following corrections to the fully symmetric expressions
of Eq.\ref{eq:total}:
\begin{eqnarray}
\label{eq:nondegeneracy}
\Delta
A_{\frac{3}{2},loop}&=&-4~ \Delta L~
A^{V}_{\frac{3}{2}}
\nonumber\\
\frac{1}{\sqrt{10}}(\Delta A_{\frac{1}{2},loop}-3~\Delta B_{loop})
&=&
- 4~ \Delta L~
\frac{1}{\sqrt{10}}(A^{V}_{\frac{1}{2}}-3B^{V})
\nonumber\\
\frac{1}{\sqrt{10}}(3~\Delta A_{\frac{1}{2},loop}+\Delta B_{loop})
& =&
6~ \Delta L~
\frac{1}{\sqrt{10}}(3A^{V}_{\frac{1}{2}}+B^{V})
\phantom{xxxxx}
\nonumber\\
\Delta
C_{loop} &=& 0
\end{eqnarray}
where
\begin{equation}
\label{eq:DeltaL}
\Delta L \equiv [L(P\rightarrow PV)-L(P\rightarrow VP^{*})]
\end{equation}
In Eq.\ref{eq:nondegeneracy} we have neglected the contribution from
weak interaction in intermediate pseudoscalar meson, as they
are much smaller than those arising from interaction in
intermediate vector meson.

{}From Eq.\ref{eq:nondegeneracy} and the fact that $\Delta L \equiv
L(P\rightarrow PV)-L(P\rightarrow VP^{*})>0$ we see that
corrections to {\bf 27}-plet amplitudes (both for   $\Delta I =
\frac{3}{2}$
and $\Delta I = \frac{1}{2}$) are negative,
and thus these amplitudes are suppressed.
On the other hand, corrections to octet $K\rightarrow \pi$ and
$K\rightarrow \eta_{8}$ amplitudes are positive and, consequently,
these amplitudes are enhanced.
The $K\rightarrow \eta_{1}$ amplitude is not modified in the
approximation under consideration.
By assuming light $\eta_{1} , \eta_{8}$ ($I=0$) mesons we slightly
overestimate hadron-level corrections to the $\Delta I~=~1/2$
amplitudes. On the other hand, the $\Delta I~=~3/2$ amplitudes
receive corrections from the $K \rightarrow \pi$ in-loop
transitions only (the $K \rightarrow \eta$ transitions change
isospin by
1/2). Thus, the estimate
of the relevant suppression factors is not affected by this
simplification.

Using
Eqs.\ref{eq:shortdistQCD},\ref{eq:isospinampl},\ref{eq:Ccoeff},
and $X^{V}/X = 13.0$ we find
from Eqs.\ref{eq:total},\ref{eq:nondegeneracy}
that
\begin{eqnarray}
\label{eq:outAmpl}
A_{\frac{3}{2}}^{out}&=&(1-52.0~\Delta L~)
A_{\frac{3}{2}}^{fact}                 \nonumber\\
A_{\frac{1}{2}}^{out}&=&(1+46.6~I+72.6~\Delta L~)
A_{\frac{1}{2}}^{fact}
\end{eqnarray}
{}From Table 1
we have $I=0.022$ and $\Delta L = 0.0093$
(for $k_{cutoff}=0.7~GeV$).
{}From Eq.\ref{eq:outAmpl} we then obtain
\begin{eqnarray}
\label{eq:enhance}
A_{\frac{3}{2}}^{out}&=& 0.52
A_{\frac{3}{2}}^{fact} \nonumber\\
A_{\frac{1}{2}}^{out}&=& 2.70
A_{\frac{1}{2}}^{fact} \nonumber\\
\frac{A_{\frac{1}{2}}^{out}}{A_{\frac{3}{2}}^{out}}&=& 5.2
\frac{A_{\frac{1}{2}}^{fact}}{A_{\frac{3}{2}}^{fact}}
\end{eqnarray}
The total hadron-level
enhancement factor of 5.2 should be compared with the number of
1.2 obtained for the case of short-distance penguin
contribution (Eq.\ref{eq:shdistpeng}).
For $k_{cutoff}=0.6,0.8~GeV$ we obtain suppression (enhancement)
factors of 0.62,0.40 (2.18,3.33)
for the {\bf 27}-plet (octet) $K\rightarrow\pi$ amplitudes
respectively.
Numerically, the hadron-level
penguin diagram enhances the octet amplitude more
than the "figure-eight" (here: $W$-exchange
\cite{Wexchange,WexchDonoghue,Wexcheng}) diagram
(the $I$ term in Eq.\ref{eq:outAmpl} is slightly larger than
the $\Delta L$ term).
The suppression of the
{\bf
27}-plet amplitudes is due to the "figure-eight" diagram. Our
numerical estimates indicate that "figure-eight" diagrams enhance
the
$\frac{A_{\frac{1}{2}}}{A_{\frac{3}{2}}}$ ratio by
a factor slightly smaller than do the penguin diagrams.
In lattice calculations "figure-eight" contributions were much smaller
than those of "eye" diagrams.
This difference between our paper and lattice calculations
seems to result from breaking
of intermediate meson degeneracy, a feature not explicitly considered
in lattice calculations.

Our estimates involve significant simplifications and cannot
be trusted to more than 50\% or so. Still, it should be
obvious that the contribution from two-meson intermediate states
is large and must be responsible for a large part of the
$\Delta I =1/2$ over $\Delta I =3/2$ enhancement providing an
overall enhancement factor of order 4-8.
Thus, long-range effects
are very important indeed.
The author hopes that,
in comparison to approaches based on the "first principles",
the estimate of these effects in hadron-level phenomenological
framework is more realistic and transparent
\cite{Blackandwhite}.

\section{Acknowledgments}
This research has been supported in part by Polish Committee for
Scientific Research Grant No. 2 P03B 231 08

\baselineskip=10pt
\renewcommand{\baselinestretch}{1.8}
\small
\normalsize
\newpage
Table 1. Dependence of $I$ and $\Delta L$ on $k_{cutoff}$ and $m_{P}$.
\\
\\
\begin{tabular}{|c||c|c|c|}
\hline
$k_{cutoff}$($GeV^{2}$) & 0.6 & 0.7  & 0.8
\\
\hline
$I(m_{P}=0.9~GeV)$  & 0.014 & 0.022 & 0.032
\\
$I(m_{P}=0.32~GeV)$ & 0.041 & 0.059 & 0.078
\\
\hline
$\Delta L$ & 0.0073 & 0.0093 & 0.0115
\\
\hline
\end{tabular}
\\
\\
\\

Table 2. Spin-flavour factors for $P\rightarrow M_{1}M_{2}$ loops
(summed over flavour)
\\
\\
\begin{tabular}{|c|c|c || l|c|c|c|c|c|c|c|}
\hline
$PV$ &$VP$ &$VV$ & & $PS$ &$PA$ &$PT$ &$VB$ &$VS$ &$VA$ &$VT$
\\
\hline
$\frac{3}{2}$ &
$\frac{3}{2}$ & $3$ & S-wave & $\frac{1}{2}$& 0& 0 &
$\frac{1}{2}$ & 0 & $1$ & 0
\\
        &   &               & D-wave & 0 & 0 & $1$ &
$1$ & 0 & $\frac{1}{2}$ & $\frac{3}{2}$ \\
\hline
\end{tabular}
\baselineskip = 10 pt
\renewcommand{\baselinestretch}{1}

\setlength {\unitlength}{1.6pt}

\begin{picture}(260,300)

\put(35,200){\begin{picture}(150,100)
\put(85,50){\line(1,0){45}}
\put(10,50){\line(1,0){45}}
\put(70,50){\oval(30,30)}
\put(10,55){$M$}
\put(120,55){$M'$}
\put(44,65){$M_{1}$}
\put(86,65){$M'_{1}$}
\put(65,24){$M_{2}$}
\put(70,65){\circle*{5}}
\put(66,75){$H_{W}^{p.c.}$}
\end{picture}}
\put(0,190){Fig.1~Weak transition in hadronic loop.}

\put(0,20){\begin{picture}(65,100)
\put(0,50){\vector(1,0){15}}
\put(15,50){\line(1,0){15}}
\put(30,50){\vector(2,1){12}}
\put(42,56){\line(2,1){12}}
\put(30,50){\vector(2,-1){12}}
\put(42,44){\line(2,-1){12}}
\put(0,55){$M$}
\put(45,70){$M_{1}$}
\put(45,25){$M_{2}$}
\put(60,48){$=$}
\end{picture}}

\put(70,20){\begin{picture}(80,100)
\put(0,55){\vector(1,0){30}}
\put(30,55){\line(2,1){40}}
\put(0,45){\line(1,0){30}}
\put(70,25){\vector(-2,1){40}}
\put(70,65){\vector(-2,-1){20}}
\put(50,55){\line(-2,-1){10}}
\put(40,50){\vector(2,-1){10}}
\put(50,45){\line(2,-1){20}}
\multiput(15,54)(0,-2.5){4}{\oval(2.5,2.5)[l]}
\multiput(60,69)(0,-2.5){4}{\oval(2.5,2.5)[l]}
\multiput(60,39)(0,-2.5){4}{\oval(2.5,2.5)[l]}
\put(5,60){$M$}
\put(60,80){$M_{1}$}
\put(60,15){$M_{2}$}
\put(73,48){$\pm$}
\end{picture}}

\put(155,20){\begin{picture}(80,100)
\put(0,55){\line(1,0){30}}
\put(70,75){\vector(-2,-1){40}}
\put(0,45){\vector(1,0){30}}
\put(30,45){\line(2,-1){40}}
\put(70,65){\line(-2,-1){20}}
\put(40,50){\vector(2,1){10}}
\put(40,50){\line(2,-1){10}}
\put(70,35){\vector(-2,1){20}}
\multiput(15,54)(0,-2.5){4}{\oval(2.5,2.5)[l]}
\multiput(60,69)(0,-2.5){4}{\oval(2.5,2.5)[l]}
\multiput(60,39)(0,-2.5){4}{\oval(2.5,2.5)[l]}
\put(5,60){$M$}
\put(60,80){$M_{1}$}
\put(60,15){$M_{2}$}
\end{picture}}

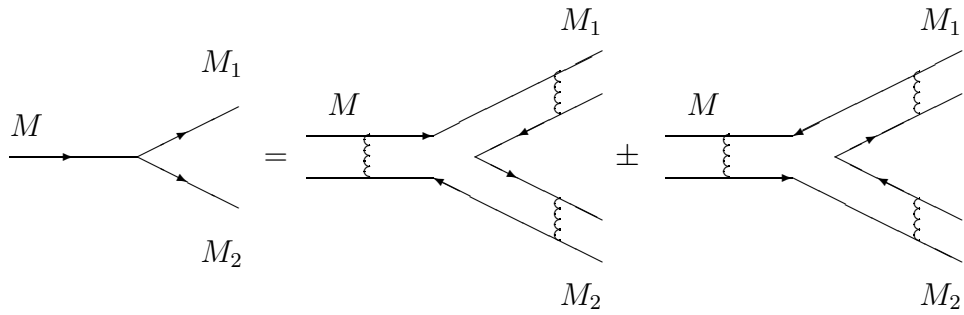
\put(0,0){Fig.2~ Diagrammatic representation of F- and D-type
strong vertices.}
\end{picture}

\newpage

\setlength {\unitlength}{1.6pt}
\begin{picture}(220,340)

\put(20,20){\begin{picture}(180,320)
\put(0,200){\begin{picture}(70,100)
\put(0,55){\vector(1,0){30}}
\put(30,55){\line(2,1){40}}
\put(0,45){\line(1,0){30}}
\put(70,25){\vector(-2,1){40}}
\put(70,65){\vector(-2,-1){20}}
\put(50,55){\line(-2,-1){10}}
\put(40,50){\vector(2,-1){10}}
\put(50,45){\line(2,-1){20}}
\multiput(15,54)(0,-2.5){4}{\oval(2.5,2.5)[l]}
\multiput(60,69)(0,-2.5){4}{\oval(2.5,2.5)[l]}
\multiput(60,39)(0,-2.5){4}{\oval(2.5,2.5)[l]}
\put(5,60){$M$}
\put(50,80){$M_{1}$}
\put(75,15){$M_{2}$}
\end{picture}}

\put(90,200){\begin{picture}(80,100)
\put(30,50){\vector(-2,1){10}}
\put(20,55){\line(-2,1){20}}
\put(30,50){\line(-2,-1){10}}
\put(0,35){\vector(2,1){20}}
\put(0,75){\vector(2,-1){40}}
\put(0,25){\line(2,1){40}}
\put(40,55){\line(1,0){30}}
\put(70,45){\vector(-1,0){30}}
\multiput(55,54)(0,-2.5){4}{\oval(2.5,2.5)[l]}
\multiput(10,69)(0,-2.5){4}{\oval(2.5,2.5)[l]}
\multiput(10,39)(0,-2.5){4}{\oval(2.5,2.5)[l]}
\put(58,60){$M'$}
\put(10,80){$M'_{1}$}
\end{picture}}

\put(70,255){\begin{picture}(20,30)
\put(0,15){\oval(10,10)[r]}
\put(20,15){\oval(10,10)[l]}
\multiput(5,15)(5,0){2}{\put(1.25,0){\oval(2.5,2.5)[t]}
\put(3.75,0){\oval(2.5,2.5)[b]}}
\multiput(0,5)(0,4){5}{\line(0,1){2}}
\multiput(20,7)(0,4){5}{\line(0,1){2}}
\multiput(2,5)(4,0){5}{\line(1,0){2}}
\multiput(0,25)(4,0){5}{\line(1,0){2}}
\end{picture}}

\put(70,215){\begin{picture}(20,30)
\put(0,10){\line(1,0){20}}
\put(0,20){\line(1,0){20}}
\end{picture}}

\put(75,190){$(a)$}


\put(0,50){\begin{picture}(70,100)
\put(0,55){\vector(1,0){30}}
\put(30,55){\line(2,1){40}}
\put(0,45){\line(1,0){30}}
\put(70,25){\vector(-2,1){40}}
\put(70,65){\vector(-2,-1){20}}
\put(50,55){\line(-2,-1){10}}
\put(40,50){\vector(2,-1){10}}
\put(50,45){\line(2,-1){20}}
\multiput(15,54)(0,-2.5){4}{\oval(2.5,2.5)[l]}
\multiput(60,69)(0,-2.5){4}{\oval(2.5,2.5)[l]}
\multiput(60,39)(0,-2.5){4}{\oval(2.5,2.5)[l]}
\put(5,60){$M$}
\put(50,80){$M_{1}$}
\put(75,15){$M_{2}$}
\end{picture}}

\put(90,50){\begin{picture}(80,100)
\put(30,50){\line(-2,1){10}}
\put(0,65){\vector(2,-1){20}}
\put(30,50){\vector(-2,-1){10}}
\put(0,35){\line(2,1){20}}
\put(0,75){\line(2,-1){40}}
\put(0,25){\vector(2,1){40}}
\put(70,55){\vector(-1,0){30}}
\put(70,45){\line(-1,0){30}}
\multiput(55,54)(0,-2.5){4}{\oval(2.5,2.5)[l]}
\multiput(10,69)(0,-2.5){4}{\oval(2.5,2.5)[l]}
\multiput(10,39)(0,-2.5){4}{\oval(2.5,2.5)[l]}
\put(58,60){$M'$}
\put(10,80){$M'_{1}$}
\end{picture}}

\put(70,105){\begin{picture}(20,30)
\put(0,15){\oval(10,10)[r]}
\put(20,15){\oval(10,10)[l]}
\multiput(5,15)(5,0){2}{\put(1.25,0){\oval(2.5,2.5)[t]}
\put(3.75,0){\oval(2.5,2.5)[b]}}
\multiput(0,5)(0,4){5}{\line(0,1){2}}
\multiput(20,7)(0,4){5}{\line(0,1){2}}
\multiput(2,5)(4,0){5}{\line(1,0){2}}
\multiput(0,25)(4,0){5}{\line(1,0){2}}
\end{picture}}

\put(70,65){\begin{picture}(20,30)
\put(0,20){\vector(2,-1){20}}
\put(20,20){\vector(-2,-1){20}}
\end{picture}}

\put(75,40){$(b)$}

\end{picture}}

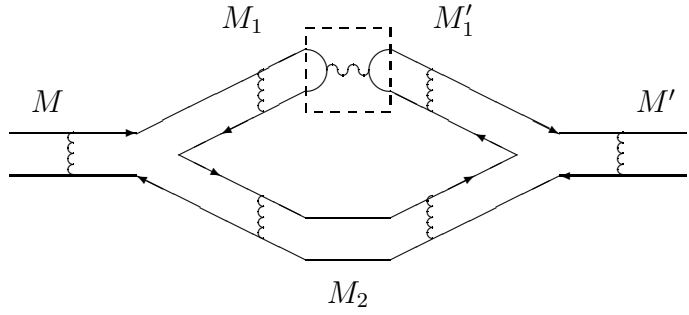
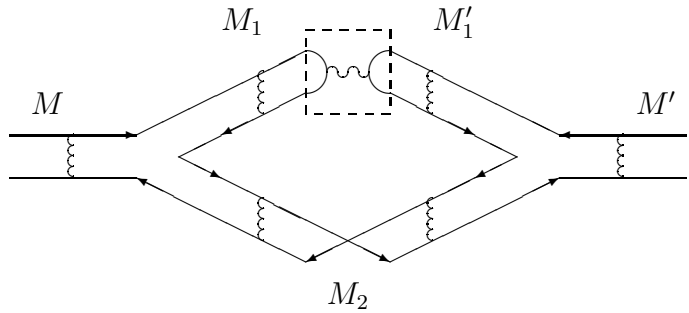
\put(0,10){Fig.3.~Quark-level diagrams generated by hadronic
loops of Fig.1:}
\put(0,0){\phantom{Fig.3.~}$(a)$~"eye" (low-energy penguin),~~
$(b)$~"figure-eight".}

\end{picture}


\begin{thebibliography}{99}
\bibitem {Blackandwhite}
In this paper penguins, though colourful inside, were built
from hadron-level colourless states. By being explicitly
black-and-white our penguins are obviously more realistic.
\end{thebibliography}
\end{document}